\documentclass{emulateapj}

\slugcomment{To appear in ApJ August 20, 2008, v683n2.}

\shorttitle{Search for Gamma-rays from Kepler's SNR}
\shortauthors{Enomoto et al.}

\begin{document}

\title{CANGAROO-III Search for Gamma Rays from Kepler's Supernova
Remnant}

\author{
R.~Enomoto,\altaffilmark{1}
Y.~Higashi,\altaffilmark{2}
T.~Yoshida,\altaffilmark{3}
T.~Tanimori,\altaffilmark{2}
G.~V.~Bicknell,\altaffilmark{4}
R.~W.~Clay,\altaffilmark{5}
P.~G.~Edwards,\altaffilmark{6}
S.~Gunji,\altaffilmark{7}
S.~Hara,\altaffilmark{8}
T.~Hara,\altaffilmark{9}
T.~Hattori,\altaffilmark{10}
S.~Hayashi,\altaffilmark{11}
Y.~Hirai,\altaffilmark{3}
K.~Inoue,\altaffilmark{7}
S.~Kabuki,\altaffilmark{2}
F.~Kajino,\altaffilmark{11}
H.~Katagiri,\altaffilmark{12}
A.~Kawachi,\altaffilmark{10}
T.~Kifune,\altaffilmark{1}
R.~Kiuchi,\altaffilmark{1}
H.~Kubo,\altaffilmark{2}
J.~Kushida,\altaffilmark{10}
Y.~Matsubara,\altaffilmark{13}
T.~Mizukami,\altaffilmark{2}
Y.~Mizumoto,\altaffilmark{14}
R.~Mizuniwa,\altaffilmark{10}
M.~Mori,\altaffilmark{1}
H.~Muraishi,\altaffilmark{15}
Y.~Muraki,\altaffilmark{13}
T.~Naito,\altaffilmark{9}
T.~Nakamori,\altaffilmark{2}
S.~Nakano,\altaffilmark{2}
D.~Nishida,\altaffilmark{2}
K.~Nishijima,\altaffilmark{10}
M.~Ohishi,\altaffilmark{1}
Y.~Sakamoto,\altaffilmark{10}
A.~Seki,\altaffilmark{10}
V.~Stamatescu,\altaffilmark{5}
T.~Suzuki,\altaffilmark{3}
D.~L.~Swaby,\altaffilmark{5}
G.~Thornton,\altaffilmark{5}
F.~Tokanai,\altaffilmark{7}
K.~Tsuchiya,\altaffilmark{16}
S.~Watanabe,\altaffilmark{2}
Y.~Yamada,\altaffilmark{11}
E.~Yamazaki,\altaffilmark{10}
S.~Yanagita,\altaffilmark{3}
T.~Yoshikoshi,\altaffilmark{1} and
Y.~Yukawa\altaffilmark{1}
}

\altaffiltext{1}{ Institute for Cosmic Ray Research, University of Tokyo, Kashiwa, Chiba 277-8582, Japan} 
\altaffiltext{2}{ Department of Physics, Kyoto University, Sakyo-ku, Kyoto 606-8502, Japan} 
\altaffiltext{3}{ Faculty of Science, Ibaraki University, Mito, Ibaraki 310-8512, Japan} 
\altaffiltext{4}{ Research School of Astronomy and Astrophysics, Australian National University, ACT 2611, Australia} 
\altaffiltext{5}{ School of Chemistry and Physics, University of Adelaide, SA 5005, Australia} 
\altaffiltext{6}{ CSIRO Australia Telescope National Facility, Narrabri, NSW 2390, Australia} 
\altaffiltext{7}{ Department of Physics, Yamagata University, Yamagata, Yamagata 990-8560, Japan} 
\altaffiltext{8}{ Ibaraki Prefectural University of Health Sciences, Ami, Ibaraki 300-0394, Japan} 
\altaffiltext{9}{ Faculty of Management Information, Yamanashi Gakuin University, Kofu, Yamanashi 400-8575, Japan} 
\altaffiltext{10}{ Department of Physics, Tokai University, Hiratsuka, Kanagawa 259-1292, Japan} 
\altaffiltext{11}{ Department of Physics, Konan University, Kobe, Hyogo 658-8501, Japan} 
\altaffiltext{12}{ Department of Physical Science, Hiroshima University, Higashi-Hiroshima, Hiroshima 739-8526, Japan} 
\altaffiltext{13}{ Solar-Terrestrial Environment Laboratory,  Nagoya University, Nagoya, Aichi 464-8602, Japan} 
\altaffiltext{14}{ National Astronomical Observatory of Japan, Mitaka, Tokyo 181-8588, Japan} 
\altaffiltext{15}{ School of Allied Health Sciences, Kitasato University, Sagamihara, Kanagawa 228-8555, Japan} 
\altaffiltext{16}{ National Research Institute of Police Science, Kashiwa, Chiba 277-0882, Japan} 

\begin{abstract}

Kepler's supernova, discovered in October 1604, produced a remnant 
that has been well studied observationally
in the radio, infrared, optical, and X-ray bands, and 
theoretically. 
Some models have predicted a TeV gamma-ray
flux that is detectable with current Imaging Cherenkov Atmospheric Telescopes. 
We report on observations carried out
in 2005 April with the CANGAROO-III telescope. 
No statistically significant
excess was observed, and limitations on 
the allowed parameter range in the model 
are discussed.

\end{abstract}


\section{Introduction}

Kepler's supernova remnant (SNR) (G4.5+6.8) 
is 400 years old \citep[see for review]{blair} and provides an 
unrivaled opportunity to verify the belief that
supernova remnants are the origin of Galactic cosmic rays.
Cas~A is younger (by $\sim$60 years) 
and was detected at TeV $\gamma$-ray energies \citep{casa}, which
implies the acceleration of high-energy cosmic rays.
Older remnants, such as
RX J0852.0-4622 \citep{katagiri,hess0852,enomoto_0852} and
RX J1713.7-3946 \citep{enomoto_nature,hess1713}, both thought to 
be 1,000$\sim$2,000 years old, have also been detected.
If SNR age was the dominant factor in for cosmic-ray acceleration,
one might expect similar levels of cosmic ray acceleration in Kepler's SNR.
Of course, other variables, such as SN type, local environment, and distance,
will certainly have some impact on the likelihood of TeV gamma-ray
detection from a SNR.

Kepler's SN was considered to be a type Ia supernova (SN) 
based on an interpretation of the
historical light curve \citep{baade}.  It was later shown that the light 
curve was also in agreement with a type II-L SN \citep{doggett}. 
However, recent observations of thermal X-ray emission by {\it ASCA} 
\citep{kinugasa} 
and {\it Chandra} \citep{reynolds} demonstrated that the SNR 
resulted from a thermonuclear supernova (type Ia), rather than the 
core-collapse of massive star (type II), even though there is  
evidence that the remnant is interacting with the progenitor star wind 
material. It may be that a type Ia event took place in a more massive 
progenitor star with a strong wind \citep{reynolds}.

\cite{ksenofontov} have modeled Kepler's SNR
and predicted a detectable TeV gamma-ray flux
under various assumptions on distances and supernova kinetic
energies that had been previously discussed in the literature.
Their prediction can be probed by the 
H.E.S.S.\,\footnote{See http://www.mpi-hd.mpg.de/htm/HESS/HESS.html}
and the future {\it GLAST}\,\footnote{See http://glast.gfsc.nasa.gov}
experiments. The CANGAROO-III
imaging Cherenkov atmospheric telescope is 
less sensitive by a factor of 3--5 than H.E.S.S.; however, 
it is able to study a specific
parameter range of the models, i.e., around the region of a supernova
explosion energy of 10$^{51}$\,erg and distance of 4.8\,kpc.
Here, we report on the result of the 2005 April observations.
As extensions to the 
\cite{ksenofontov} theory have successfully explained 
the gamma-ray fluxes from
other historical SNRs, it is important to investigate and
constrain the allowed parameter ranges in this model
with measurements of fluxes or upper limits.  

\section{CANGAROO-III Stereoscopic System}

The CANGAROO-III stereoscopic system consists of four imaging atmospheric
Cherenkov telescopes located near Woomera, South Australia (31$^\circ$S,
137$^\circ$E).
Each telescope has a 10\,m diameter segmented reflector, 
consisting of 114 spherical mirrors 
made of fiber-reinforced plastic \citep{kawachi}, each of 80\,cm diameter,
mounted on a parabolic
frame with a focal length of 8\,m.
The total light-collecting area is 57.3\,m$^2$.
The first telescope, T1, which was the CANGAROO-II telescope
\citep{enomoto_nature},
is not presently in use due to its smaller field of view
and higher energy threshold.
The second, third, and fourth telescopes (T2, T3, and T4) were operated for the
observations described here.
The camera systems for T2, T3, and T4 are identical and 
are described in \citet{kabuki}.
The telescopes are located at the 
eastern (T1), western (T2), southern (T3) and northern (T4)
corners of a diamond 
with sides of $\sim$100\,m \citep{enomoto_app}.
The point-spread functions of these telescopes 
are 0.$^\circ$24.

\section{Observations}

The observations were carried out 
during the period from 2005 April 11 to 17 (UT)
using the ``wobble mode"
in which the pointing position of each telescope was
shifted in declination by $\pm$0.5 degree 
every 20 minutes \citep{wobble}
from the target:
(RA, dec [J2000]) = (262.$^\circ$671, $-$21.$^\circ$486).  
We made no OFF source runs, as the wobble mode enables OFF-source
regions to be observed simultaneously with the target regions.
The sensitive region in wobble mode observations
is considered as being within one degree
from the average pointing position.
This SNR is located 6.$^\circ$8 from the Galactic plane; therefore,
no significant diffuse gamma-ray background is 
expected within the field of view.

In the observations, the hardware trigger used to select any two telescope
hits was employed \citep{nishijima}.
The images in two out of three telescopes were required to have clusters
of at least five adjacent pixels exceeding a 5\,photoelectron threshold
(off-line two-fold coincidence).
To illustrate the effect of this criterion, 
the event rate was reduced from 10$\sim$12 to 
6$\sim$8\,Hz for T3--T4 coincidences, 
depending on the elevation angle.
Looking at the time dependence of these rates, we can remove data
taken under cloudy conditions. 
The effective observation time was
874 minutes, and the mean zenith angle was 15.$^\circ$2.

The light-collecting efficiencies, including the reflectivity
of the segmented mirrors, the light guides, and the quantum efficiencies
of the photomultiplier tubes were monitored by a muon-ring analysis
\citep{enomoto_vela} with individual trigger data during the
same period. 
The average light yield per unit arc-length of muon rings 
is approximately proportional
to the light-collecting efficiencies.
Deterioration in these efficiencies is mostly due to dirt and dust 
settling on the mirrors and light guides, which are
washed annually to improve their reflectivities.
In analyzing T2 data, we had some difficulties in detecting muon-rings
during this period; therefore, we did not use T2 in this analysis.
Unfortunately these observations were made shortly before 
regular mirror washing.
Also we had some mis-setting of 
the T2 
ADC-gate width in this period.
This analysis, therefore, used only T3 and T4 
two-fold coincidence data.

\section{Analysis}

The analysis procedures used here were identical to those 
described in \citet{cena}, except for the point that these were two-fold
coincidence data.
More details can be found in \citet{enomoto_vela} and
\citet{enomoto_0852}.
Here, we briefly describe them.

At first, the image moments of $Width$ and $Length$ \citep{hillas} 
were calculated for the two
 telescopes.
The incident direction of the gamma-ray was determined by minimizing
the sum of the squared widths (weighted by the photon yield) 
of the two images seen from the assumed position (fitting parameter)
with a constraint on the distances from the intersection point to each
image center.

In order to derive the gamma-ray likeliness,
we used  
the Fisher Discriminant (hereafter $FD$) \citep{fisher,enomoto_vela}.
The input parameters were
$$\vec{P}=(W3,W4,L3,L4),$$
where $W3,W4,L3,L4$ are energy-corrected $Widths$ and $Lengths$ for 
T3 and T4.

We rejected events with any hits in the outermost layer of the cameras
(``edge cut"). These rejected events were potentially 
incompletely sampled, resulting in errors particularly
in the $Length$ distribution, which would have produced 
deformations of the $FD$. 

Then $FD$ distributions were derived on a position-by-position basis.
Comparing those in the signal region and the control background
region with the Monte-Carlo expectation, 
we can derive the number of gamma-ray--like events.
Here, we assume the $FD$ distribution of the gamma-ray signal to
be that derived from Monte-Carlo simulations.
In the gamma-ray simulations we used a spectrum 
proportional to $E^\gamma$, where $\gamma$=$-$2.1$\pm$0.2.
Fits of the $FD$ distribution of the source position
with the above simulated
signal and control background functions were carried out 
to derive the number of gamma-ray--like events.
This was a one-parameter fitting with the constraint that
the sum of the signal and the background events 
corresponds to the total number of events, i.e.,
the fitted parameter can be derived exactly analytically.

\section{Results}

Since the spatial size of Kepler's SNR (100$"$) is much less than our
angular resolution ($\sigma$ = 0.24$^\circ$), 
we concentrate here on searching for
a point source near the target center.

\begin{figure}
\plotone{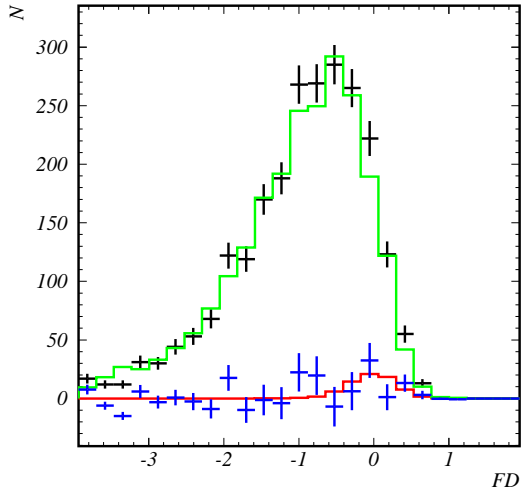}
\caption{
$FD$ distribution for the region inside $\theta^2~<~0.06$ deg$^2$.
The black points with error bars are those for the above region,
the green histogram is for $\theta^2$ inside the (0.1--0.2) [deg$^2$]
region,
the blue points with error bars are subtracted data using 
the results of the fit, that is,
``gamma-ray--like" events, and the red histogram is the best-fit gamma-ray,
i.e., the gamma ray response function.
}
\label{fig1}
\end{figure}

In order to determine whether or not there is a
gamma-ray excess around the SNR,
we made the
$FD$ distribution within the PSF ($\theta^2<0.06\simeq 0.24^2$)
and fitted it with 
a background function derived
from the $FD$ distribution in the region $\theta^2$=(0.1--0.2), 
and a signal function from Monte-Carlo simulations.
The fitting parameter is the ratio of gamma-rays to total number of events.
The fitting results are
shown in Fig. \ref{fig1}.
The best-fit excess was 71$\pm$32 events,
where the uncertainty is the 1$\sigma$ statistical error.
The gamma-ray response function from the Monte-Carlo
simulation is shown by the red histogram.
The threshold of this analysis is estimated
from the Monte-Carlo simulation to be $\sim$500\,GeV.
The systematic error on the energy determination
 is considered to be less than 15\%, with
the main factor being the
uncertainty in the light collection efficiency and atmospheric
conditions.

\begin{figure}
\plotone{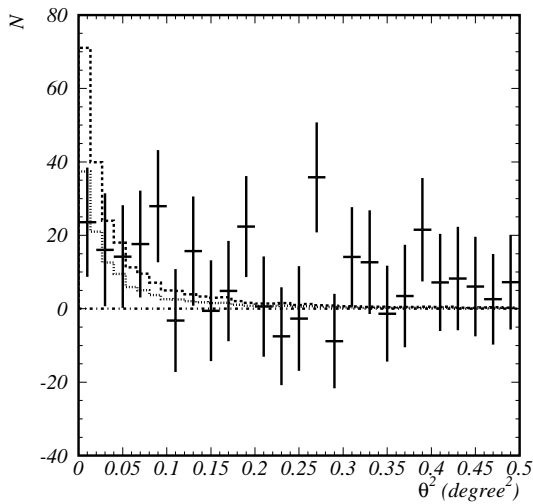}
\caption{$\theta^2$ plot in the unit of deg$^2$.
The points with error bars are the result of fit, that is,
the ``gamma-ray--like" event
distribution as a function of $\theta^2$.
The dot-dashed line is the zero level.
The (light) dotted histogram is the best-fit for our point-spread function.
The (heavy) dashed histogram is the two-$\sigma$ upper limit for the
point-source assumption.}
\label{fig2}
\end{figure}

We then made a radial distribution of gamma-ray--like events.
$FD$ distributions in various $\theta^2$ slices were made.
The control background region was again selected in the $\theta^2$ range of
between 0.1 and 0.2 deg$^2$. 
The standard fitting procedure described above was carried out.
The fitted result is shown in Fig.\ \ref{fig2}.
The reduced $\chi^2$ for a null assumption (the dot-dashed line) 
is $\chi^2$/DOF\,=\,25.4/25 (where DOF is degrees of freedom).
The best fit with the point spread function (PSF) is shown by the
dashed histogram, where $\chi^2$/D.O.F\,=\,18.1/25.
The dashed histogram is the 2$\sigma$ upper limit
(135-event excess) for the PSF excess.

\begin{figure}
\plotone{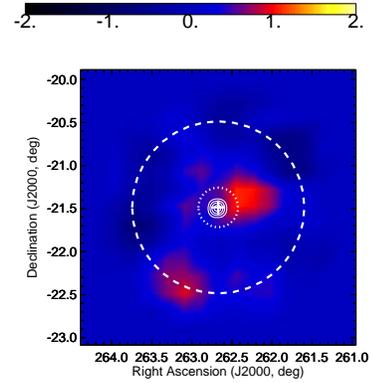}
\caption{
Significance map. 
The average telescope pointing position is indicated by the
white cross at the center.
The dotted-white circle is the point-spread 
function and the searched region. The dashed-white circle is the fiducial
region (1$^\circ$ radius). The thin-white contours are
the radio measurement at 4850\,MHz \citep{skyview} which are well inside
the searched region.
The color bar indicates the excess in standard deviations
over the background (see text for details).
}
\label{fig3}
\end{figure}

In order to examine the morphology, we segmented the field of view into
0.2\,$\times$\,0.2 degree$^2$ square bins. The $FD$ distributions
for corresponding bins were made and fitted. The control-background
region is defined as
the second-closest layer of 16 bins, 
all of which are more than 0.3\,deg from the center of the target region,
i.e., larger than the 0.24 degree point-spread function (PSF).
The statistics of the
control-background are, therefore, sixteen times larger than that
of the signal bin.
The results are shown in Fig.~\ref{fig3}.
We smoothed the results by averaging the neighboring nine bins.
Our sensitivity falls off significantly beyond
one degree in radius from the center. The
PSF is the 0.24 degree radius circle 
which fully contains
the radio observed SNR (the thin-white contours).
According to the Monte-Carlo simulations, 65\% of gamma-rays
from this SNR should be contained in this circle.
The PSF is not a Gaussian and has a broader tail component.
In order to contain 90\% of events, we need to broaden this cut
to 0.5 degree, resulting in a loss of sensitivity. 
We, therefore, selected
the cut at 0.24 degree (1$\sigma$ region).
The significance distributions (excess divided by the statistical error
before smoothing)
are approximately normal (Gaussian) distributions with
a mean value of 0.21$\pm$0.13,
and a standard deviation of 1.19$\pm$0.12,
consistent with null assumption within systematic uncertainties.
The statistical significance of
the maximum located 0.$^\circ$35 west-north-west from the center
is 
(before smoothing)
3.3$\sigma$, and therefore not compelling.

There is no statistically-significant excess that is consistent
with emission from Kepler's SNR convolved with the telescope PSF.
The derived upper limits (ULs) for
the gamma-ray flux are shown in Table~\ref{table1}.
\begin{table}
\caption{The 2$\sigma$ upper limits to the integral fluxes from
Kepler's SNR at five energy thresholds. 
The spectral
index, $\gamma$, of the energy spectrum (E$^{-\gamma}$) is that used in the 
Monte-Carlo gamma-ray simulations. Note that the gamma-ray acceptance
depends on $\gamma$.}
\label{table1}
\begin{center}
\begin{tabular}{ccc}
\hline\hline
$\gamma$ & Threshold [GeV]& Upper Limits [cm$^{-2}$s$^{-1}$]\\
\hline
2.1 & 530 & 1.7 $\times$ 10$^{-11}$ \\
2.1 & 680 & 1.1 $\times$ 10$^{-11}$ \\
2.1 & 930 & 6.8 $\times$ 10$^{-12}$ \\
2.1 &1300 & 1.5 $\times$ 10$^{-12}$ \\
2.1 &2400 & 6.2 $\times$ 10$^{-13}$ \\
1.9 & 550 & 1.7 $\times$ 10$^{-11}$ \\
1.9 & 700 & 1.2 $\times$ 10$^{-11}$ \\
1.9 & 930 & 7.5 $\times$ 10$^{-12}$ \\
1.9 &1300 & 1.6 $\times$ 10$^{-12}$ \\
1.9 &2400 & 7.7 $\times$ 10$^{-13}$ \\
2.3 & 510 & 1.8 $\times$ 10$^{-11}$ \\
2.3 & 650 & 1.2 $\times$ 10$^{-11}$ \\
2.3 & 930 & 6.5 $\times$ 10$^{-12}$ \\
2.3 &1200 & 1.5 $\times$ 10$^{-12}$ \\
2.3 &2300 & 5.6 $\times$ 10$^{-13}$ \\
\hline\hline
\end{tabular}
\end{center}
\end{table}
Here, we used a $E^{-2.1\pm 0.2}$ spectrum for the gamma-ray
simulation.
The ULs range between 10--30\% of the Crab nebula flux. 
The statistically insignificant
excess near the center of the field of view,
shown in Figs.~\ref{fig1}, \ref{fig2}, and \ref{fig3} only 
appeared in the lower
energy regions. At higher energies,
we do not see any excess.
Therefore the ULs at lower energies were higher than that at higher energies.

\section{Discussion}

\begin{figure}
\plotone{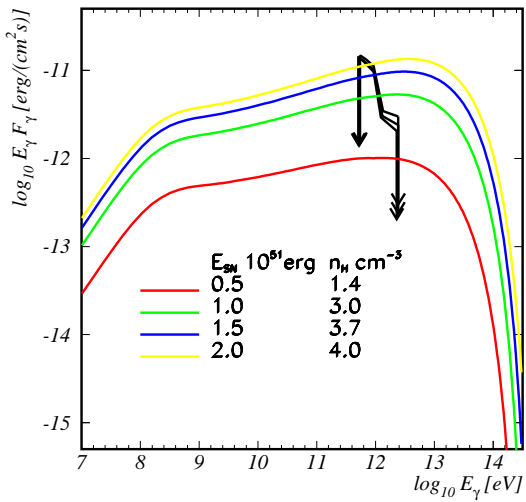}
\caption{
Spectral energy distributions for a fixed distance of 4.8\,kpc. 
The black curves with arrows
at both ends are the upper limits obtained by these observations.
Three curves are obtained by different acceptance corrections
using three types of energy spectra in the Monte-Carlo
simulation (see the caption of Table~\ref{table1}).
The colored curves are the theoretical predictions, which are
as same as Fig.~3 of \cite{ksenofontov}.
}
\label{fig4}
\end{figure}

\begin{figure}
\plotone{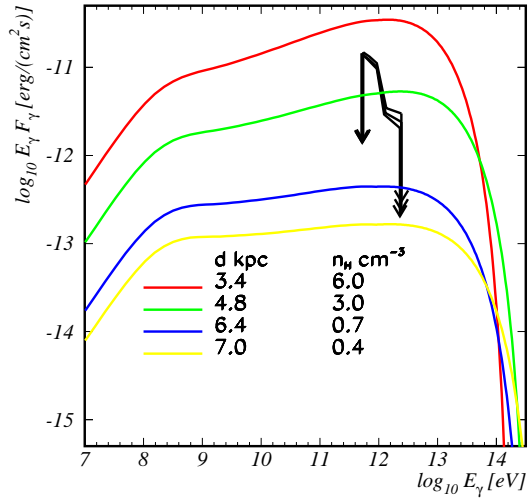}
\caption{
Spectral energy distributions
for a fixed supernova explosion energy 
of 10$^{51}$\,erg. The black curve with arrows
at both ends are the upper limits obtained by this observation.
Three curves are obtained by different acceptance corrections
using three types of energy spectra in the Monte-Carlo
simulation (Table~\ref{table1}).
The colored curves are the theoretical predictions, which are
as same as Fig.~4 of \cite{ksenofontov}.
}
\label{fig5}
\end{figure}

The upper limits given in Table \ref{table1} are plotted on 
spectral energy distributions in 
Figs. \ref{fig4}  and \ref{fig5}.
The vertical and horizontal units were fitted to Figs.~3 and 4 in 
\cite{ksenofontov} in order to discuss the allowed parameter
ranges with respect to our observational upper limits.
This theory considered a reasonably wide range of possibilities for
the distance of this SNR (3.4--7\,kpc) and 
the supernova explosion energy (0.5--2\,$\times$\,10$^{51}$\,erg).
Other adopted parameters, such as the cosmic ray injection rate, 
expansion rate, and electron-to-proton ratio, while plausible, are open to
debate. We do not review those details here but refer readers to
the discussion in \cite{ksenofontov}.
The black curve with arrows at both ends were 
obtained from this observation.
Three types of the energy spectra
($\propto~E^\gamma : \gamma=-1.9,~-2.1,~-2.3$)
 were used for acceptance
correction using the Monte-Carlo simulation.
The uncertainty due to the assumption of the energy spectral index is
small on a logarithmic scale.
The colored curves were obtained from a theoretical prediction
by \cite{ksenofontov}.
In Fig.~\ref{fig4} the distance to the object was
fixed to be 4.8 kpc.
The red, green, blue, and yellow curves correspond to different
supernova explosion kinetic energies and corresponding number densities of
ambient circumstellar material, which came from fitting to the
observed shock radius and speed.
The most probable is the second one 
(the green curve) and our upper limits
are (in part) below it, meaning that 
for a distance of 4.8\,kpc the explosion energy 
should be less than $\sim10^{51}$\,erg. 
Although the green curve is close to a best estimation, a
large allowable range of parameter space remains.

Fig.~\ref{fig5} is the case when the supernova explosion energy 
is fixed at 10$^{51}$\,erg, and several distances 
in the allowed range \citep{reynoso} are assumed.
Distances less than $\sim$5\,kpc are (for this fixed SN energy) not favored, 
suggesting that this SNR
is marginally more distant than the best current observational estimations.
Of course, this conclusion also assumes that all other assumptions 
in the model are
correct, for example, that the expansion is in Sedov phase, that the 
supernova was Type Ia, that 10\% of kinetic energy is transferred to the
cosmic-ray energy, etc.

\begin{figure}
\plotone{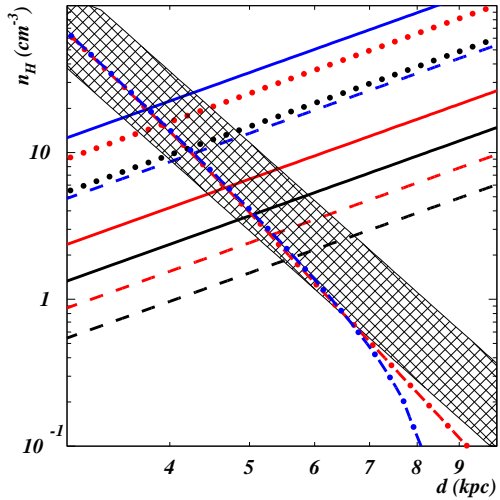}
\caption{
The allowed region in the plane of the distance ($d$)
versus ambient density ($n_{\rm H}$), based
on the neutral pion decay $\gamma$-ray emission model where we assume 
that the total number spectrum of protons is proportional to a power law 
with an exponential cutoff and the value of the conversion efficiency from 
the explosion energy to the cosmic-ray energy is 10\%.
The solid, dashed, and dotted lines indicate the upper limits obtained from 
the observed 2$\sigma$ upper limits in Table \ref{table1}, compared with 
the integral fluxes of the models for the power-law indices 2.1, 1.9, and 2.3, 
respectively. The colors of the lines represent the cutoff energy of protons; 
the black, red, and blue ones are obtained with 
the values $10^{15}$, $10^{14}$, and 
$10^{13}$ eV, respectively.
The allowed regions lie under these upper-limit lines.
The hatched region indicates the region which satisfies the 
Sedov-Taylor solution of the apparent radius of the Kepler's SNR, 100$"$ at 
400 yr with the explosion energy $0.5\sim2\times 10^{51}$ erg. 
The dot-dashed lines are obtained from the approximate analytic blast-wave 
positions of the radius 100$"$, assuming 
the solution of \cite{TrueloveMcKee}.
The red and blue ones are obtained 
assuming the ejecta power-law index $n=6$ and $n=14$, 
respectively,
with ejecta mass of 1.4$M_{\sun}$ with the explosion 
energy $10^{51}$ erg. 
}
\label{fig6}
\end{figure}

The constraints on the parameters of the distance $d$ and the ambient
density $n_{\rm H}$ are illustrated in Fig. \ref{fig6}. 
Here we assume that the total number spectrum of protons is proportional 
to a power law with an exponential cutoff $E^{-p}\exp(-E/E_{max})$ 
and that the neutral pion decay $\gamma$-ray emission dominates. 
If the conversion efficiency ($\epsilon$) 
from the explosion energy $E_{\rm sn}$ to 
the cosmic-ray energy is assumed to be 10\%
(i.e., $\epsilon=0.1$), the normalization factor 
of the proton spectrum can be determined and the $\gamma$-ray fluxes 
can be calculated \citep{Mori} on the assumption of the power-law index 
$p$ and the cutoff energy $E_{\rm max}$. 
Given the parameters of the $E_{\rm sn}$, $p$, and $E_{\rm max}$, 
the upper limits of $n_{\rm H}/d\,^2$ are calculated from the observed 
2$\sigma$ upper limits in Table~\ref{table1}, compared with the integral 
fluxes of the model, because the $\gamma$-ray fluxes 
$F_{\gamma}$ are proportional to $\epsilon E_{\rm sn} n_{\rm H}/d\,^2$. 
In Fig.~\ref{fig6}, we plotted the upper limits 
for the power-law indices 2.1 (solid), 1.9 (dashed), 
and 2.3 (dotted) and for the cutoff energies of protons $10^{15}$
(black), $10^{14}$ (red), 
and $10^{13}$ eV (blue). We did not plot the case of $p=2.3$ and
$E_{\rm max}=10^{13}$\,eV, as the flux in the GeV energy region 
exceeds the EGRET upper limit. 

The apparent radius $\theta$=100$"$ of the Kepler's SNR at $t_{\rm
age}$ gives another constraint.  Here we consider two solutions on the 
expansion law of the blast-wave shock. The first one is the Sedov-Taylor 
solution : 
$\theta d \propto (E _{\rm sn}/n_{\rm H})^{1/5}  t_{\rm age}^{2/5}$. 
Another one is the approximate analytic solutions \citep{TrueloveMcKee}, 
which can be applied from the ejecta-dominated phase to the 
Sedov-Taylor phase.  In the latter case, the extra parameters
of ejecta mass $M_{\rm ej}$ 
and the ejecta power-law index $n$, are added to the explosion energy 
$E _{\rm sn}$ and the ambient matter number density $n_{\rm H}$. 
The region which is satisfied with the Sedov-Taylor 
solution of the apparent radius of the Kepler's SNR, 100$"$ at 400 yr and 
$E_{\rm sn}=0.5\sim2\times 10^{51}$ erg are shown as the hatched 
one in Fig.\,6. 
The dot-dashed lines are obtained from the approximate analytic 
solution with ejecta mass of 1.4$M_{\sun}$ with the explosion energy
$10^{51}$ erg with two kinds of the ejecta power-law index $n=6$ and $n=14$.
In the case of $p=2.1$ and $E_{\rm max}=10^{14}$ eV, we note
that the CANGAROO-III upper limit implies that the Kepler's SNR is located
at a distance larger than about 4.5\,kpc. For the maximum upper limit in 
the case of $p=2.1$ and $E_{\rm max}=10^{13}$ eV, this means
that the SNR is located at a distance larger than about 3.6\,kpc.

The distance of a type Ia supernova can be estimated using the correlation
between the shape of the optical light curves and the intrinsic luminosity
of SNe. Before knowledge of this correlation, \cite{baade} studied the
historical light curve of the Kepler's SN and classified it as a type Ia,
and \cite{danziger} estimated the distance of $3.2 \pm 0.7$ kpc
using only the maximum luminosity. We can now fit Baade's data with the
improved light-curve template of a type Ia \citep{jha}: $d=4.0 \pm
0.4$ kpc on the assumption of a visual extinction $3.27 \pm 0.14$ mag
\citep{schaefer}, although the fitted light curve after 100 days is
not a good fit.
 
This value seems to be marginally consistent with our lower limit
obtained by this TeV $\gamma$-ray observations. On the other hand, 
based on the study of H\,{\sc i} kinematics and the association of 
H\,{\sc i} cloud with 
the SNR, \cite{reynoso} put a lower limit of $d=4.8 \pm 1.4$\,kpc and an upper
limit of 6.4\,kpc on the distance. These estimations do not contradict our
lower limit.

In any case, a part of the plausible region in parameter space 
has been rejected, although a large allowed range remains. 
Although we did not detect any signal in these 15 hours of observations, 
future detections may strongly constrain the theory.
As found in Figs.\,3 \& 4 in
\cite{ksenofontov}, the sensitivity of H.E.S.S.\ is much
lower than the theoretical predictions.
The future GLAST mission will also enable the lower energy
range to be probed.
Future large Cerenkov telescope arrays, such as CTA
\footnote{See http://www.mpi-hd.mpg.de/htm/CTA/}
will allow even more sensitive observations to be made.

\section{Conclusion}

TeV gamma-ray observations toward
the 400 year old remnant of Kepler's SN 
were made in 2005 April.
Although a measurable
flux of TeV-gamma rays had been predicted, 
we did not observe any statistically significant
excess in this region, and the constraints on 
the allowed parameter range have been discussed.
Although a region of parameter range has been rejected, 
more sensitive measurements are required to
constrain the models further.

\acknowledgments

We thank Dr. L.T.\ Ksenofontov for various discussions on the 
estimated gamma-ray flux from Kepler's SNR.
We thank  Dr. N.\ Yasuda for discussions on the distance of this SNR.
This work was supported by a Grant-in-Aid for Scientific Research by
the Japan Ministry of Education, Culture, Sports, Science and Technology, 
the Australian Research Council, JSPS Research Fellowships,
and Inter-University Researches Program 
by the Institute for Cosmic Ray Research.
We thank the Defense Support Center Woomera and BAE Systems.

\end{document}